# The electronic structure of the doped one-dimensional transition metal oxide $Y_{1-x}Ca_xBaNiO_5$ studied using x-ray absorption

Z. Hu, M. Knupfer, M. Kielwein, U. K. Rößler, M. S. Golden, and J. Fink,

Institute for Solid State Research, IFW Dresden Dresden, D-01171 Dresden, Germany

F.M.F. de Groot

Department of Inorganic Chemistry, University of Utrecht

Sorbonnelaan 16, 3584 CA Utrecht, The Netherlands

T. Ito

Electrotechnical Laboratory, 1-1-4 Tsukuba, Ibaraki 890-5788, Japan

G. Kaindl

Institut für Experimentalphysik, Freie Universität Berlin, Arnimallee 14, D-14195 Berlin, Germany

## Abstract

A strong anisotropic distribution of the holes in Ni 3d and O 2p orbitals is observed in the polarization dependent O 1s and Ni $2p_{3/2}$ x-ray absorption spectroscopy of the linear-chain nickelate $Y_{1-x}Ca_xBaNiO_5$ (x = 0, 0.05, 0.1, 0.2), which demonstrates the one-dimensional nature of the electronic state in these compounds. Furthermore, the additional holes introduced by Ca-doping occupy both O 2p and Ni 3d orbitals along the $NiO_5$ chains. By comparing the experimental Ni $2p_{3/2}$ absorption spectra of $Y_{1-x}Ca_xBaNiO_5$ to those from charge transfer multiplet calculations we can derive the orbital character of the additional holes to be of ~60% O2p and ~40% Ni 3d.

**I. Introduction**

The fundamental physical properties of the onedimensional (1D) transition metal oxides have attracted a lot of interest in recent years [1-3]. For instance, the discovery of superconductivity in $Sr_{0.4}Ca_{1.36}Cu_{24}O_{41+\delta}$, which has no $CuO_2$ plane, but rather a linear chain of $CuO_4$ plaquettes and a two-leg ladder network, has evoked a series studies of the quasi-1D cuprates [5-8]. Furthermore, a detailed theoretical description can be much easier achieved in the case of one-dimensional systems as compared to their two- or three-dimensional relatives, and thus comprehensive studies of one-dimensional highly correlated systems can help to develop or test different model approaches. One of the fascinating aspects of 1D systems lies in Haldane's conjecture that the behavior of antiferromagnetic (AF) quantum spin chains depends crucially on whether the spin is integer or half-integer [9]. The hole doped 1D system $Y_{2-x}Ca_xBaNiO_5$ discovered recently [10] is an ideal system to explore this conjecture. Neutron scattering data have indicated that unlike 2D transition metal (TM) oxides and 1D cuprates [5-8], $Y_2BaNiO_5$ is not an antiferromagnet but a compound with a quantum disordered ground state and a Haldane gap [3]. This is consistent with the expectation that a $3d^8$ configuration (S = 1) exhibits a gap $\Delta_m$ between a collective spin ground state and antiferromagnetic order. Neutron scattering studies of $Y_2BaNiO_5$ showed that $\Delta_m/J = 0.3{\sim}0.4$ (J being the spin coupling within the chain) [11,12], which is in agreement with the prediction $\Delta_m/J \sim 0.41$ from various calculations [13-17]. Recent band structure calculations proposed that substitutional doping should inhibit the AF behavior and produce a metallic system [18]. A combined study of polycrystalline $Y_{2-x}Ca_xBaNiO_5$ using photoemission and bremsstrahlung isochromat spectroscopy (BIS) showed that the hole doping only slightly reduces the occupied part of the electronic structure and more pronounced changes are observed for the unoccupied part [19]. The modification of the unoccupied electronic states upon Ca-doping in $Y_{2-x}Ca_xBaNiO_5$ can also be probed using polarization-dependent x-ray absorption spectroscopy (XAS) on single crystals. This technique allows one to measure the unoccupied states both site and orbital selectively and has a significantly higher experimental resolution than BIS measurements do. Previously, O 1s and Ni $2p_{3/2}$ XAS spectra of $Y_{2-x}Ca_xBaNiO_5$ were reported for a substitution level of x = 0.15 [3]. In this work, we report a more systematic study of $Y_{2-x}Ca_xBaNiO_5$ with x = 0, 0.05, 0.1, 0.2 using polarization dependent XAS measurements of O 1s and Ni $2p_{3/2}$ core excitations. In addition, we compare our experimental results to theoretical simulations which allows the study of the distribution of the doping induced holes in between the O 2p and Ni 3d orbitals.

## II. Experimental

$Y_{2-x}Ca_xBaNiO_5$ single crystals were grown using the traveling-solvent floating zone technique. $Y_{2-x}Ca_xBaNiO_5$ adopts an orthorhombic crystal structure [20]. As shown in Fig. 1, distorted corner-sharing $NiO_6$ octahedra form one-dimensional chains along the crystallographic **a**-axis of this system. The Ni-O-Ni angle is 180º and the Ni-O(1) and Ni-O(2) bond lengths are 1.88 Å and 2.18 Å, respectively. Furthermore, the octahedra are slightly squeezed in **b**-direction as compared to the **c**-direction, i.e. the local symmetry of Ni in $Y_{2-x}Ca_xBaNiO_5$ is lowered to $D_{2h}$. Samples with flat **ac** and **ab** surfaces of 3x3 mm$^2$ size were prepared using an ultra-microtome equipped with a diamond knife. The XAS experiments were carried out in the non-surface-sensitive fluorescence yield detection mode using linearly polarized light from the SX700-II monochromator operated by the Freie Universität Berlin at BESSY I. The energy resolution of the monochromator was set to be 280 meV and 600 meV at the O 1s and Ni-2p thresholds, respectively. The measurements were carried out with a normal incidence of the synchrotron radiation. The resulting data were normalized at 80 eV above the absorption thresholds. A correction for self-absorption effects was performed according to a procedure described elsewhere [21,22]. In order to ease a direct comparison of our results with those from other Ni-O based compounds we label the relevant O and Ni orbitals with z along the crystallographic **a** direction (i.e. along the Ni-O chains) and with x/y perpendicular to this direction.

**III. Results and discussions**

In Fig. 2 we show the O 1s XAS spectra of $Y_{2-x}Ca_xBaNiO_5$ for the light polarization vector **E** parallel to the crystallographic **a** (filled circles) and **b** (open circles) axes. These measurements probe unoccupied electronic states with O $2p_z$ (**E**||a) and O $2p_x$ (**E**||b) character. The energetically lowest lying structure in the spectra of $Y_2BaNiO_5$ which is labeled A in Fig. 2 is found at 531.7 eV for **E**||a and at 531.4 eV for **E**||b. These energies are close to that observed for the corresponding spectral feature of NiO and these structures are thus assigned to transitions from the O 1s level into unoccupied states with O 2p character mixed with the Ni 3d-derived so-called upper Hubbard band (UHB). The shift of this feature to higher energies for **E**||a compared to **E**||b is attributed to the distortion of the $NiO_6$ octahedra along the **a** axis which results in differing band widths in the two directions and different on-site energies for the corresponding oxygen sites. This interpretation is in agreement with predictions from band structure calculations [18]. The larger intensity for **E**||a compared to **E**||b reflects the stronger covalence between the O 2p and Ni 3d levels, i.e. a larger contribution of O 2p character to the UHB along the **a** axis (chain) with a shorter Ni-O1 distance.

At a low doping level of x = 0.05, a weak pre-edge peak B appears at 528.3 eV for **E**||a and at 528.4 eV for **E**||b, originating from transitions into the Ca-doping induced O 2p hole states. Upon further Ca-doping, the intensity of peak B increases more strongly for **E**||a, which clearly indicates that the doping induced holes mainly occupy orbitals along the chain direction, thus underlining the one-dimensionality of the low lying electronic structure of $Y_{2-x}Ca_xBaNiO_5$. At a doping level of x = 0.1 a weak feature C is observed at 529.6 eV for **E**||a,

similar to what has been found for x = 0.15 in a previous study [3]. Furthermore, similar spectral features were also observed in the case of $La_{2-x}Sr_xNiO_4$, where a weak structure corresponding to C was more clearly observed for higher Sr doping levels ($\geq 0.4$) [23].

Formally, the introduction of holes leads to the formation of trivalent Ni(III) sites with a $3d^7$ configuration. Magnetic susceptibility measurements indicate the formation of spin S=1/2 sites upon Ca-doping [24]. As the covalence between Ni and O switches on, the Hamiltonian matrix and the ground state of a $3d^n$ system are given by:

$$H = \begin{vmatrix} 0 & T \\ T & \Delta \end{vmatrix} \quad (1)$$

$$\Phi_g = \alpha_0 |3d^n> + \beta_0 |3d^{n+1}\underline{L}>, \quad (\alpha_0^2+\beta_0^2) = 1 \quad (2)$$

$$\beta_0/\alpha_0 = \{(\Delta^2 + 4T^2)^{1/2} - \Delta\} / 2T \quad (3)$$

where n = 7 stands for Ni(III) and $\underline{L}$ denotes a hole in O 2p states (for simplicity we neglect any admixture of $d^{n+m}\underline{L}^m$ configurations with $m \geq 2$). $\Delta$ and T in Eq. (1) and (3) are the charge transfer energy between Ni 3d and O 2p orbitals and the corresponding transfer integral. Thus, the $3d^7$ state is mixed with $3d^8\underline{L}$ configurations and the degree of this covalent mixture is represented by $\beta_0$. Since only intra-atomic transitions are allowed in an XAS experiment, the intensity of peaks B and C in Fig. 2 is proportional to $\beta_0^2$. There are three different possible $|3d^8\underline{L}>$ configurations which can mix with $|3d^7>$. These are depicted in Fig. 1b. Two of the $|3d^8\underline{L}>$ configurations involve a hybridization of O $2p_z$ orbitals with Ni $3d(3z^2-r^2)$ states denoted as (II) with spin S = 1 and (III) with S = 0, while the third one consists of a hybrid of Ni 3d(xy) with and O $2p_x/2p_y$ states denoted as (IV) with S=0. The two states with O $2p_z$ contributions can be experimentally observed in our set up for **E**∥**a**. As a consequence of Hund's rule, the low spin $3d^8$ states (III) have a higher energy and therefore are more weakly populated as compared to the high spin $3d^8$ configuration (II). The theoretical prediction for this energy splitting between the low and high spin Ni $3d^8$ configurations is 1.3. eV [17] which is comparable with the observed energy separation between peak B and C for **E**||**a** in Fig. 2. We thus assign peaks B and C for **E**||**a** to excitations into high and low spin Ni $3d^8$ states, respectively, which are hybridized with O $2p_z$ orbitals pointing along the one-dimensional chains in $Y_{2-x}Ca_xBaNiO_5$. The broad pre-edge peak for **E**||**b** consequently originates from the hybridized state resulting from the mixture between the $3d_{xy}$ and the $O2p_x/O2p_y$ states as shown schematically in Fig. 1b(IV). Its excitation energy relative to the corresponding peaks for **E**||**a** is determined by the Hund's rule coupling as well as by the different on-site energies of the O $2p_x/2p_y$ and the O $2p_z$ sites resulting from different Madelung potentials. The significantly higher spectral weight of the pre-edge features for **E**||**a** demonstrates that the Ni-O hybridization is stronger along the $NiO_5$ chains than perpendicular to them which is a direct consequence of the shorter Ni-O bond length along the chains. Finally, at the bottom of Fig. 2 we directly compare the spectra for **E**||**a** of the compounds with x = 0 (solid line) and x = 0.2. It is clearly seen that the spectral weight of peak A is reduced upon doping which is a result of a spectral weight transfer form the UHB (peak A) to the doping induced features B and C. Such a spectral weight transfer is a direct signature of a strongly correlated electron system (here a charge transfer insulator) and has been studied experimentally [25,26] and theoretically [27.28] for other correlated materials.

Figure 3 shows the Ni-$L_3$ XAS spectra of $Y_{2-x}Ca_xBaNiO_5$ and $Nd_{0.9}Sr_{1.1}NiO_4$ as a 2D Ni(III) reference compound. The spectral profile of $Y_2BaNiO_5$ for **E**||**b** (open circles) consists of a dominant peak A at 853 eV and a broad shoulder B at about 2 eV higher energy. The spectral profile for **E**||**c** is very similar to that for **E**||**b** and is not showed here. In contrast to the O 1s spectra, the main feature A is less intensive for **E**||**a** than for **E**||**c**. This confirms the above interpretation that there is a larger O 2p contribution to the UHB along the **a** axis.

In addition, for **E**∥**a** the dominant peak A is shifted by 0.7 eV to higher energy with respect to that for **E**∥**b** while the high energy shoulder B appears at the same energy (855 eV) in both cases. Since the spectra for **E**∥**a** and **E**∥**b** result from transitions from Ni $2p_{3/2}$ to Ni $3d(3z^2-r^2)$ and $3d(xy)$ states, respectively, the energy shift of peak A of about 0.7 eV is basically a measure for the crystal field splitting between $3d(3z^2-r^2)$ and $3d(xy)$ states caused by the pseudo-tetragonal distortion of the $NiO_6$ octahedra in $Y_2BaNiO_5$. Thus, the tetragonal splitting of the $e_g$-derived 3d states is about half of the octahedral $e_g$-$t_{2g}$ crystal field splitting (10 Dq) which has been estimated to be 1.3-1.5 eV for the related three-dimensional compound NiO [29,30].

We note that the tetragonal splitting observed here for $Y_2BaNiO_5$ is much smaller than that previously reported for $La_2NiO_{4-\delta}$ of 1.3 eV [23]. We argue that in the latter case the reported experimental value is overestimated, as it results from the analysis of the Ni $2p_{1/2}$ absorption spectrum where the higher energy shoulder strongly changes in intensity as a function of polarization, rending the analysis of the tetragonal splitting difficult and less conclusive. The corresponding Ni $2p_{3/2}$ absorption spectrum of $La_2NiO_{4-\delta}$ cannot be evaluated as it overlaps in energy with La 5d-related absorption features.

Upon Ca-doping the spectral weight of the shoulder B increases more strongly in the case of **E**∥**a** compared to **E**∥**b**. This again demonstrates that the additional holes in the $NiO_5$ chains of $Y_2BaNiO_5$ predominantly occupy orbitals that are oriented along the chain axis. A Ca-doping induced increase of the higher energy shoulder (B) mainly for **E**∥**a** was also reported previously for a doping level of x = 0.15 [3]. Generally, this increase of spectral weight at the high energy side upon doping parallels the changes seen in the Ni $2p_{3/2}$ XAS of the compound $Nd_{2-x}Sr_xNiO_{4-\delta}$ with increasing Sr doping, and indicates the transition from a formally divalent to a formally trivalent Ni-O system [31]. To further illustrate this, we have included the Ni $2p_{3/2}$ spectrum of the formally trivalent two dimentional (2D) nickel compound $Nd_{0.9}Sr_{1.1}NiO_4$.

We note that in the latter case, there is no single crystal available. This, however, does not change the conclusions drawn below as $Nd_{0.9}Sr_{1.1}NiO_4$ contains essentially undistorted $NiO_6$ octahedra, i.e. the XAS spectra are essentially polarization independent. The spectral features of $Nd_{0.9}Sr_{1.1}NiO_4$ are similar to those observed in an XAS study of other formally trivalent Ni compounds, such as $PrNiO_3$ and $NdNiO_3$ [32]. Fig. 3 demonstrates that upon doping, i.e. with increase of the formal valence from Ni(II) to Ni(III), spectral weight is transferred from the lower energy peak A to peak B, which reflects an increase of the effective Ni 2p core hole potential due to poorer screening resulting from decrease in the valence electron count. Furthermore, the $Ni-2p_{3/2}$ spectra become broader due to an increase of the O 2p admixture in the wave function, since $\Delta$ is expected to decreased by as much as 3 eV going from Ni(II) to Ni(III) compounds [31,33], as result of the increase of $\beta_0$ according to Eq. (3).

The polarization dependent Ni $2p_{3/2}$ XAS spectra of undoped $Y_2BaNiO_5$ can be well reproduced using charge-transfer atomic multiplet calculations applying using (ionic) tetragonal crystal-field values 10 Dq = 0.8 eV and Ds = 0.2 eV [31,33]. The $3d^9\underline{L}$ state is found to lie 3 eV above the $3d^8$ state in the ground state of a Ni(II) compound, whereby the charge transfer energy $\Delta$ is defined here as the energy difference between the energetically lowest $3d^n$ and the lowest $3d^{n+1}\underline{L}$ state. The solid and the dashed curves below the experimental data for $Y_2BaNiO_5$ and $Nd_{0.9}Sr_{1.1}NiO_{3.95}$ in Fig. 3 depict the theoretical results for **E**∥**a** and **E**∥**b**, respectively. The ground state consists of a mixture of 85% $3d^8$ and 15% $3d^9\underline{L}$ which reflects the predominantly ionic character of the Ni(II) compounds. The theoretical spectrum of the Ni(III) compound is reproduced by decreasing the charge transfer energy from 3 eV for $Y_2BaNiO_5$ to 0.4 eV for $Nd_{0.9}Sr_{1.1}NiO_{3.95}$. Consequently, the $3d^7$ weight in the ground state of $Nd_{0.9}Sr_{1.1}NiO_{3.95}$ was found to be 40%, which is much smaller than 85% $3d^8$ contribution in $Y_2BaNiO_5$ (Ni[II]) indicating a significant increase of covalence on going from Ni(II) to Ni(III) in agreement with previous studies of other Ni-O materials [31].

For a small doping level x ≤ 0.2, the $3d^7$ weight is certainly smaller than that in the reference compound $Nd_{1.1}Sr_{0.9}NiO_{3.95}$, but one cannot conclude that the doping induced holes have exclusively mainly O 2p character. In order to illustrate the distribution of the additional holes between the Ni and O orbitals involved, we show in Fig. 4 a comparison of the Ni $2p_{3/2}$ (top of Fig. 4) and the O 1s (bottom of Fig. 4) spectra taken at doping levels x = 0 and x = 0.2 and for **E**||**a**, considering the fact shown above that the doping induced holes are predominantly occupying orbitals along the $NiO_5$ chains. Going from x = 0 to x = 0.2, we observe an increase of intensity by 12% and 8% for a sum of the O1s features A, B and C and the Ni $2p_{3/2}$ features A and B, respectively. This would suggest that the doping induced holes have about 40% Ni 3d character. The Ni-O-Ni bond angle of 180º present along the $NiO_5$ chains favors significant O 2p character for the additional holes as compared to an isolated $NiO_6$ cluster via the inter-octahedra hybridization in the ground state of the former. It was found previously that the covalence was decreased (the $3d^7$ weight was increased) by nearly 10% going from compounds with interconnected to systems with isolated $NiO_6$ clusters [31]. In addition, this delocalization of the electronic states along the chains offers the system a further opportunity to screen the Ni 2p core hole via ligand electrons in such a core level spectroscopy. As a result, the corresponding spectrum becomes broader and the higher energy spectral weight is suppressed due to a final state effects. A more quantitative analysis of the distribution of the additional holes in the doped compounds based upon theoretical simulation would be intensively, but is beyond the scope of the present paper.

**IV. Summary**

We have presented systematic O 1s and Ni $2p_{3/2}$ x-ray absorption spectroscopy studies of the doped one-dimensional transition metal oxide $Y_{1-x}Ca_xBaNiO_5$ (x = 0, 0.05, 0.1, 0.2). Our results demonstrate that the doping induced holes predominantly occupy O 2p and Ni 3d orbitals that are oriented along the $NiO_5$ chains in this compound. This emphasizes the one-dimensional nature of the electronic states in this material. The energy splitting between a high and low spin Ni $3d^8$ configuration could be experimentally determined to be about 1.3 eV in good agreement with theoretical predictions. A comparison of the Ni $2p_{3/2}$ absorption spectra to those from a formally trivalent nickel oxide as well as to cluster calculations showed that the doping induced holes occupy both O 2p and Ni 3d orbitals, with the ratio for x = 0.2 being roughly 2/3.

**Acknowledgement:**

We acknowledge financial support by the Deutsche Forschungsgemeinschaft (SPP1073 [(Fi439/7-1, Fi439/10-1] and SFB 463 "Seltenerd- Übergangsmetallverbindungen: Struktur, Magnetismus und Transport").

**Figure captions**

Fig. 1. Schematic view of (a) the $NiO_6$ clusters which form the $NiO_5$ chains in $Y_2BaNiO_5$ and (b) the $3d^7$ and $3d^8\underline{L}$ configurations discussed in the text.

Fig. 2. O 1s x-ray absorption spectra of $Y_{2-x}Ca_xBaNiO_5$ for the light polarization vector **E** parallel to the crystal axes (filled circles **E**||**a**; open circles **E**||**b**). At the bottom the spectrum for **E**||**a** and x = 0 is shown again as solid line for comparison.

Fig. 3. Ni $2p_{3/2}$ x-ray absorption spectra of $Y_{2-x}Ca_xBaNiO_5$ for the light polarization vector **E** parallel to the crystal axes (filled circles **E**||**a**; open circles **E**||**b**) as well as of $Nd_{0.9}Sr_{1.1}NiO_{3.95}$ as a Ni(III) reference compound. The theoretical curves obtained carrying out charge transfer multiplet calculations are given as solid/dashed lines below the data.

Fig. 4. Comparising the Ni-$2p_{3/2}$ (top) and O 1s (bottom) x-ray absorption spectra between doping levels of x = 0 and x = 0.2 for **E**||**a**.

(a)

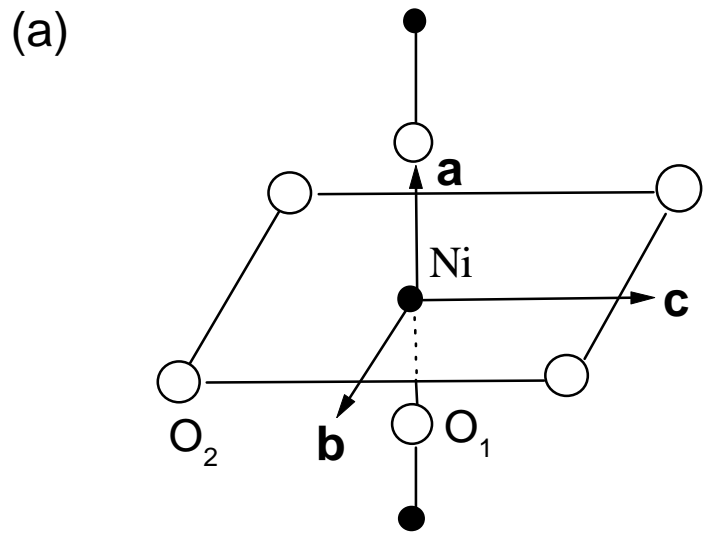

(b)

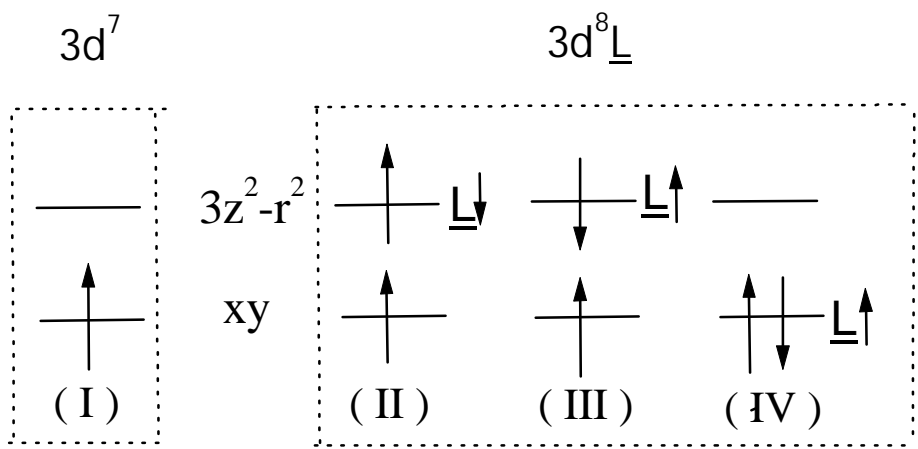

Fig. 1

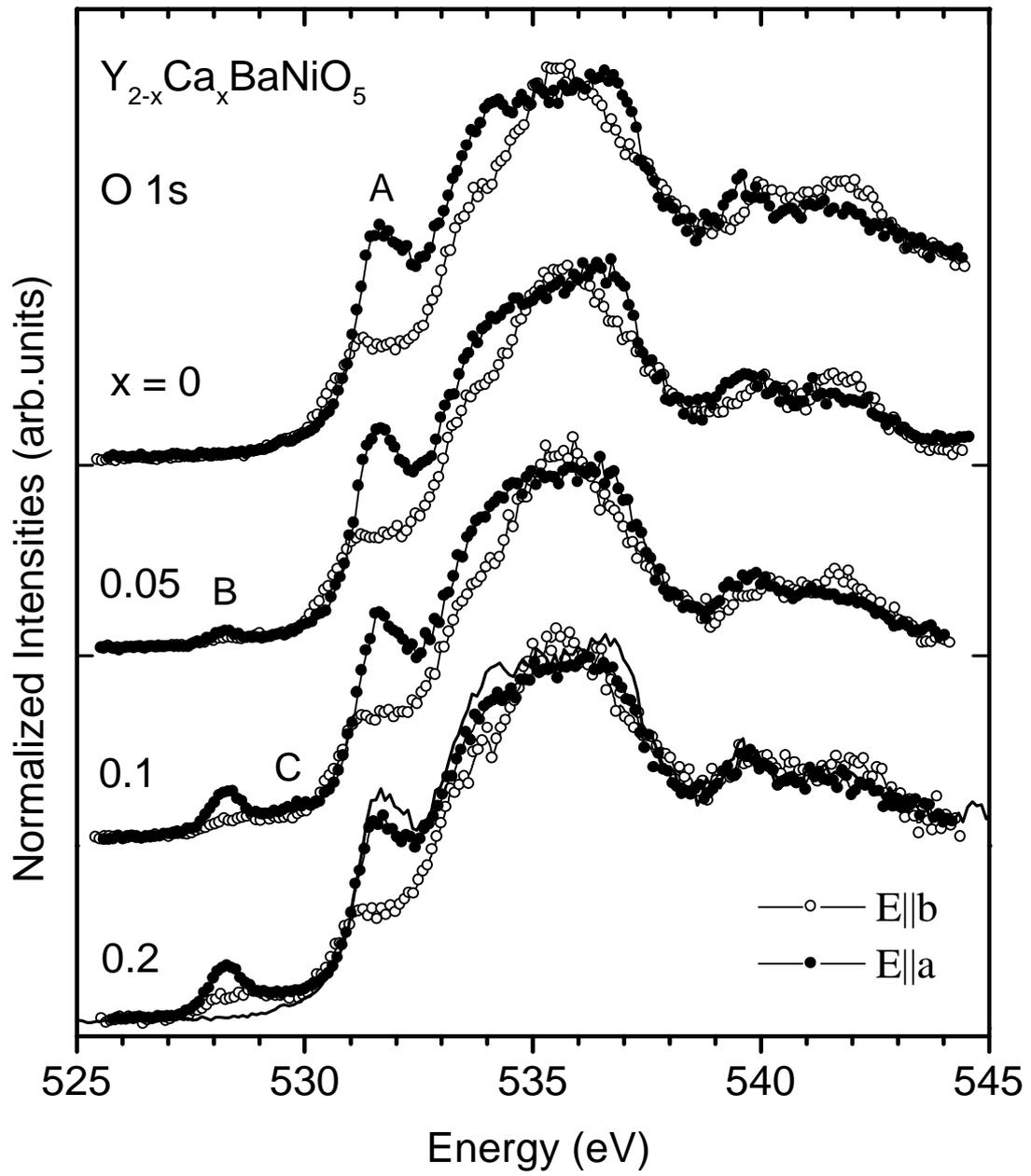

Fig. 2    Z. Hu *et al.*

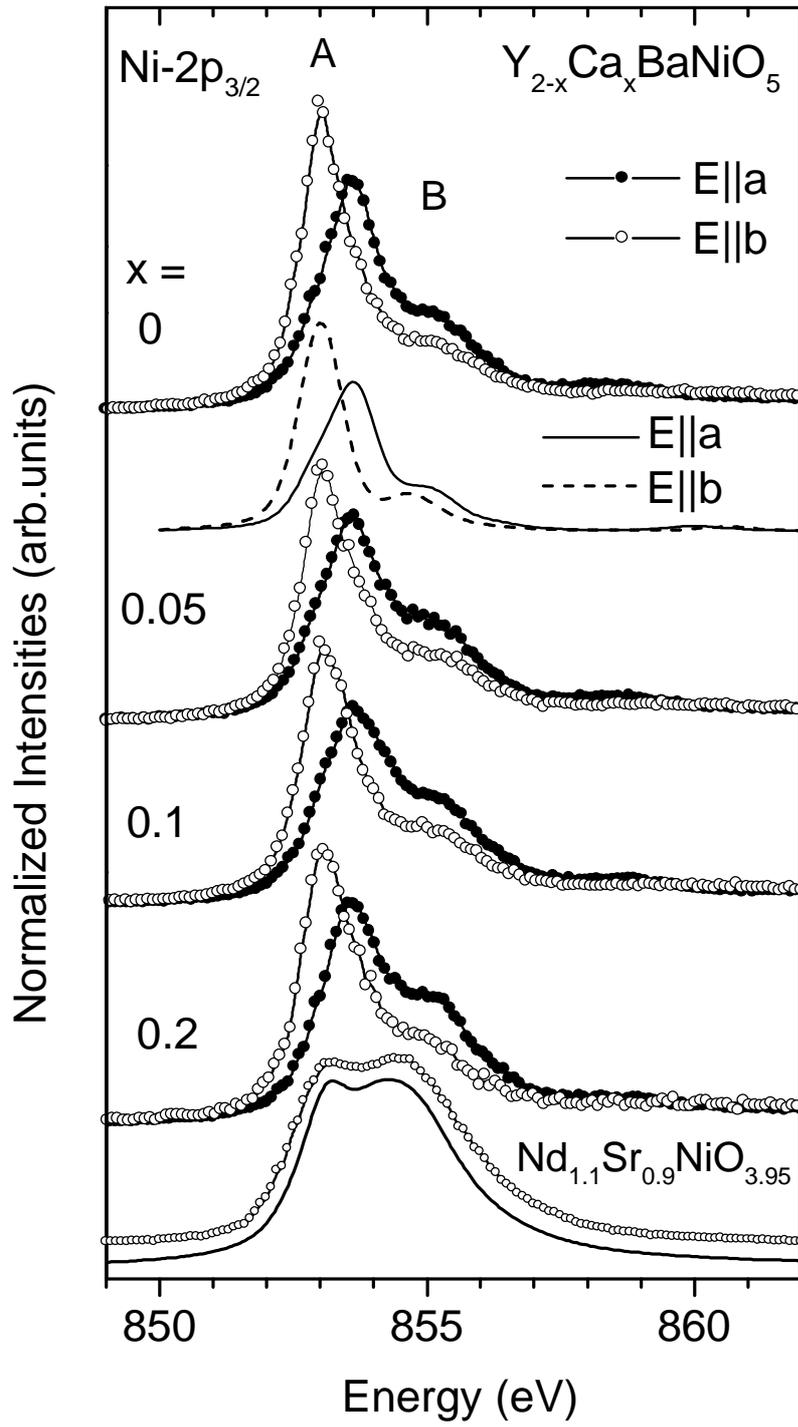

Hu et al., Fig. 3

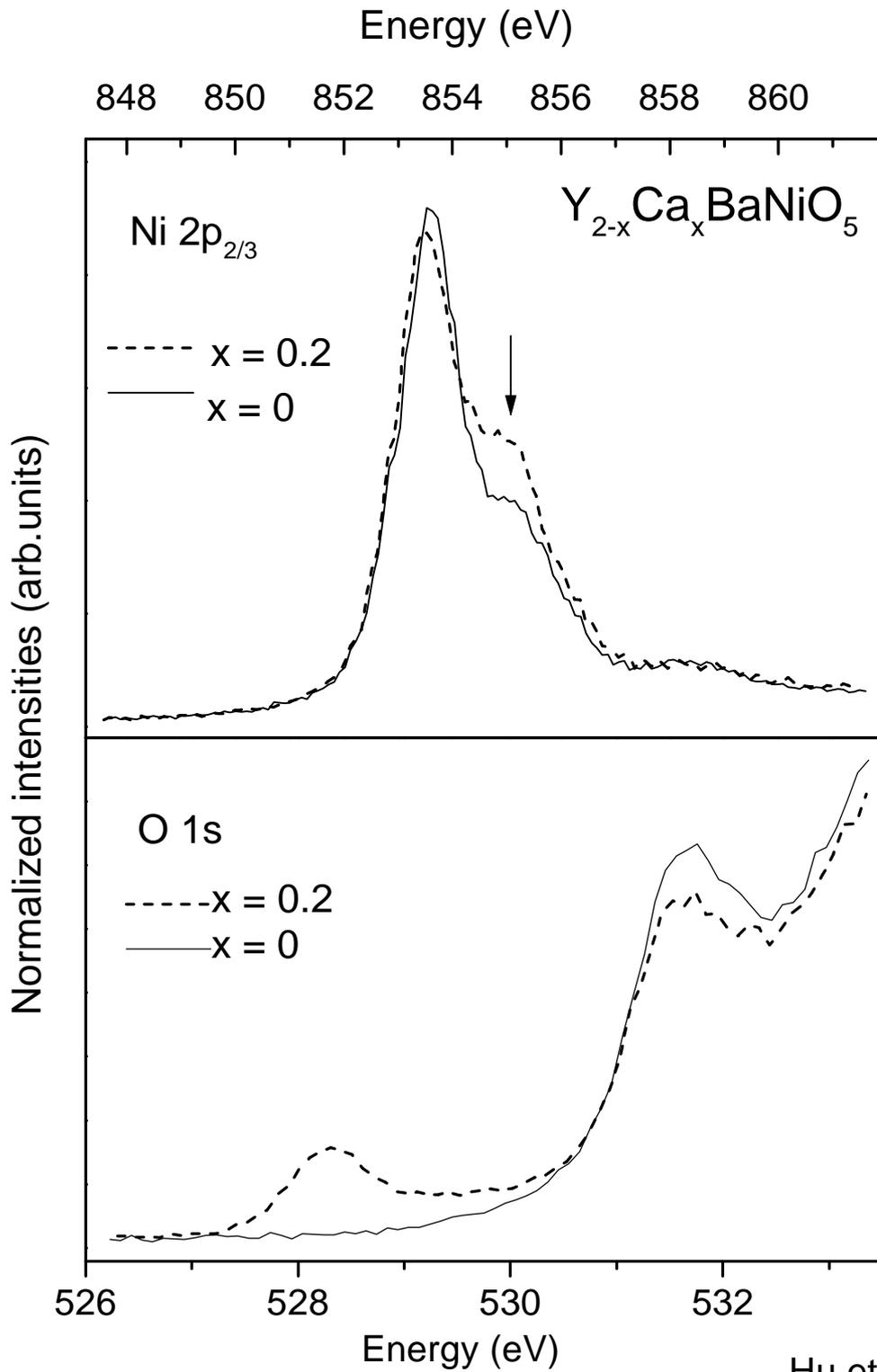

Hu et al., Fig. 4